
\magnification=1200
\baselineskip 13pt

\def\Ht{${\cal H}_{tot}$}
\def\At{${\cal A}_{tot}$}
\font\meinfont=cmbx10 scaled \magstep2
\centerline{\meinfont FROM QUANTUM PROBABILITIES}\par
\centerline{\meinfont TO CLASSICAL FACTS}\vskip 2 truecm
\settabs\+~Faculty of Physics and BiBoS~~~~~~~~~~~~~~~~~~
&Institute of Theoretical Physics &\cr
\+~~~~~~~Ph. Blanchard &A. Jadczyk &\cr
\+~~~~~~~Faculty of Physics and BiBoS
&Institute of Theoretical Physics &\cr
\+~~~~~~~University of Bielefeld &University of Wroclaw &\cr
\+~~~~~~~D-33615 Bielefeld &Pl. Maxa Borna 9 &\cr
\+~ &PL-50204 Wroclaw &\cr
\vskip 1 truecm
\centerline{a Gianfausto Dell'Antonio con stima ed amicizia}
\vskip 3 truecm
\centerline{Abstract}\medskip
Model interactions between classical and quantum systems are
briefly reviewed. These include: general measurement - like
couplings, Stern-Gerlach experiment, model of a counter,
quantum Zeno effect, piecewise deterministic Markov processes and
meaning of the wave function.
\vfill\eject
\noindent\underbar{\bf 1. Introduction}\medskip
Quantum theory seems to be in conflict with so much of the real
world but agrees so perfectly well with observations and describes
and computes so much from the behaviour of the solids, the colour of the
stars up to the structure and the function of DNA that almost all
physicists accepted the fascinating but hard to follow interpretation
of its mathematical structure proposed by N. Bohr [1]. As emphasized
by F. Rohrlich [2] ``Quantum Theory is the greatest conceptual
revolution of our century and probably the greatest that mankind
had ever experienced ...''. In classical statistical mechanics we have
to do with ontic determinism but epistemic indeterminism. In Quantum
Mechanics the probabilistic description \underbar{is} the
fundamental description and no deeper level exists i.e. we have
ontic indeterminism (what events are possible and how possible each of
them is). The quantum state contains a lot of potentialities and the
collapse creates Heisenberg's transition from the possible to the
actual. Orthodox Quantum Mechanics considers two types of incompatible
time evolution $U$ and $R$, $R$ denoting the reduction of the
quantum state.
$U$ is linear, deterministic, local,
continuous and time reversal invariant. On the other hand $R$ is
probabilistic, non-linear, discontinuous, anticausal (making events
consequences of their observations and determines a stochastic change
of reality in time).
For a fundamental physical
theory this situation is not very satisfactory but it works for all
practical purposes!\par
In recent papers [3], [4] we propose a mathematically consistent
model describing the interaction between classical and quantum
systems. It provides an answer to the question of how and why quantum
phenomena become real as a result of interaction between quantum and
classical domains. Our results show that a simple dissipative time
evolution can allow a dynamical exchange of information between
classical and quantum levels of Nature. Indeterminism is an implicit
part of classical physics. Irreversible laws are fundamental and
reversibility is an approximation. R. Haag formulated the same thesis
as ``... once one accepts indetermination there is no reason against
including irreversibility as part of the fundamental laws of Nature''
[5].\par
With a properly chosen initial state the quantum probabilities are
exactly mirrored by the state of the classical system and moreover
the state of the quantum subsystem converges as $t \to + \infty$
to a limit in agreement with von Neumann-L\"uders standard quantum
mechanical measurement projection postulate $R$. In our model the
quantum system $\Sigma_q$ is coupled to a classical recording
device $\Sigma_c$ which will respond to its actual state. $\Sigma_q$
should affect $\Sigma_c$, which should therefore be treated dynamically.
We thus give a minimal mathematical semantics to describe the
measurement process in Quantum Mechanics. For this reason the simplest
model that we proposed can be seen as the elementary building block
used by Nature in the constant communications that take place between
the quantum and classical levels. We propose to consider the total
system $\Sigma_{tot} = \Sigma_q \otimes \Sigma_c$ and the behaviour
associated to the total algebra of observables ${\cal Q}_{tot} =
{\cal Q}_q \otimes {\cal Q}_c = {\cal C} (X_c) \otimes {\cal L}
({\cal H}_q)$, where $X_c$ is the classical phase space and
${\cal H}_q$ the Hilbert space associated to $\Sigma_q$, is now
taken as the fundamental reality with pure quantum behaviour as an
approximation valid in the cases when recording effects can be
neglected. In ${\cal Q}_{tot}$ we can describe irreversible changes
occuring in the physical world, like the blackening photographic
emulsion, as well as idealized reversible pure quantum
and pure classical processes.
We extend the model of Quantum Theory in such a way that the successful
features of the existing theory are retained but the transitions
between equilibria in the sense of recording effects is permitted.
In Section 2 we will briefly describe the mathematical and physical
ingredients of the model and discuss the measurement process in this
framework.\par
The range of applications of the model is rather wide as will be
shown in Section 3 with a discussion of Zeno effect. To the Liouville
equation describing the time evolution of statistical states of
$\Sigma_{tot}$ we will be in position to associate a piecewise
deterministic process taking values in the set of pure states of
$\Sigma_{tot}$. Knowing this process one can answer all kinds of
questions about time correlations of the events as well as simulate
numerically the possible histories of individual quantum-classical
systems. Let us
emphasize that nothing more can be expected from a theory without
introducing some explicit dynamics of hidden variables. What we
achieved is the maximum of what can be achieved, which is more than
orthodox interpretation gives. There are also no paradoxes; we cannot
predict, but we can simulate the observations of individual systems.
Moreover, we will briefly comment on the meaning of the wave
function. Section 4 deals with some other applications and concluding
remarks.
\vskip 1.5 truecm
\noindent\underbar{\bf 2. Measurement-like processes}\medskip
For a long time the theory of measurement in quantum mechanics,
elaborated by Bohr, Heisenberg und von Neumann in the 1930s has
been considered as an esoteric subject of little relevance for real
physics. But in the 1980s the technology has made possible to transform
``Gedankenexperimente'' of the 1930s into real experiments. This
progress implies that the measurement process in quantum theory is now
a central tool for physicists testing experimentally by high-sensitivity
measuring devices the more esoteric aspects of Quantum Theory.\par
Quantum mechanical measurement brings together a macroscopic and a
quantum system.
\vskip 1 truecm
\noindent{\underbar{\bf 2.1 Interacting classical and quantum systems}}
\medskip
Let us briefly describe the mathematical framework we will use. A
good deal more can be said and we refer the reader to [3,4]. Our aim
is to describe a non-trivial interaction between a quantum system
$\sum_q$ in interaction with a classical system $\sum_c$.
To the quantum system there corresponds a Hilbert space ${\cal H}_q$.
In ${\cal H}_q$ we consider a family of orthonormal projectors
$e_i = e_i^* = e_i^2,~(i = 1,...,n),~\sum^n_{i=1} e_i = 1$,
associated to an observable $A = \sum^n_{i=1} \lambda_i e_i$ of the
quantum mechanical system. The classical system is supposed to have
$m$ distinct pure states, and it is convenient to take $m \ge n$.
The algebra ${\cal A}_c$ of classical observables is in this case
nothing else as ${\cal A}_c = {\bf C}^m$. The set of classical states
coincides with the space of probability measures. Using the notation
$X_c = \{s_0, ..., s_{m-1})$, a classical state is therefore an
$m$-tuple $p = (p_0, ..., p_{m-1}), p_\alpha \ge 0,~\sum^{m-1}_{\alpha
= 0}~p_\alpha = 1$. The state $s_0$ plays in some cases a distinguished
role and can be viewed as the neutral initial state of a counter. The
algebra of observables of the total system ${\cal A}_{tot}$ is given
by
$$
{\cal A}_{tot} = {\cal A}_c \otimes L ({\cal H}_q) = {\bf C}^m
\otimes L ({\cal H}_q) = \bigoplus^{m-1}_{\alpha=0}~
L ({\cal H}_q),\eqno(1)
$$
and it is convenient to realize ${\cal A}_{tot}$ as an algebra of
operators on an auxiliary Hilbert space ${\cal H}_{tot} =
{\cal H}_q \otimes {\bf C}^m = \bigoplus^{m-1}_{\alpha=0}~{\cal H}_q$.
${\cal A}_{tot}$ is then isomorphic to the algebra of block diagonal
$m \times m$ matrices $A = diag(a_0, a_1, ..., a_{m-1})$ with
$a_\alpha \in L({\cal H}_q)$. States on \At~are represented by block
diagonal matrices
$$
\rho = diag(\rho_0, \rho_1, ..., \rho_{m-1})\eqno(2)
$$
where the $\rho_\alpha$ are positive trace class operators in
$L({\cal H}_q)$ satisfying moreover $\sum_\alpha~Tr(\rho_\alpha) = 1$.
By
taking partial traces each state $\rho$ projects on a `quantum state'
$\pi_q (\rho)$ and a `classical state' $\pi_c (\rho)$ given
respectively by
$$
\pi_q (\rho) = \sum_\alpha \rho_\alpha,\eqno(3)
$$
$$
\pi_c (\rho) = (Tr \rho_0, Tr \rho_1, ..., Tr \rho_{m-1}).\eqno(4)
$$
The time evolution of the total system is given by a semi group
$\alpha^t = e^{tL}$ of positive maps\footnote{$^1$}{In fact, the maps
we use happen to be also completely positive.} of \At -- preserving
hermiticity, identity and positivity -- with $L$ of the form
$$
L(A) = i [H,A] + \sum^n_{i=1} (V_i^* A V_i - {1 \over 2} \{V_i^* V_i,
A\}).\eqno(5)
$$
The $V_i$ can be arbitrary linear operators in $L$(\Ht) such that
$\sum V_i^* V_i \in$ \At~and $\sum V_i^* A V_i$\hfill\break
$\in$ \At~whenever
$A \in$ \At, $H$ is an arbitrary block-diagonal self adjoint operator
$H = diag (H_\alpha)$ in \Ht~and $\{ , \}$ denotes anticommutator
i.e.
$$
\{ A,B \} \equiv AB + BA.\eqno(6)
$$
In order to couple the given quantum observable $A = \sum^n_{i=1}
\lambda_i e_i$ to the classical system, the $V_i$ are chosen as
tensor products $V_i = \sqrt \kappa e_i \otimes \phi_i$,
where $\phi_i$
act as
transformations on classical (pure) states. Denoting $\rho (t) =
\alpha_t (\rho (0))$, the time evolution of the states is given by
the dual Liouville equation
$$
\dot \rho (t) = -i [H, \rho (t)] + \sum^n_{i=1} (V_i \rho (t)
V_i^* - {1 \over 2} \{V_i^* V_i, \rho (t)\}),\eqno(7)
$$
where in general $H$ and the $V_i$ can explicitly depend on time.
\hfill\break
\leftline{\bf Remarks:}\hfill\break
1) It is possible to generalize this framework for the case where
the quantum mechanical observable $A$ we consider has a continuous
spectrum (as for instance in a measurement of the position) with
$A = \int_{\bf R} \lambda dE (\lambda)$. See [6,7] for more details.
\hfill\break
2) Since the center of the total algebra \At~is invariant under
any automorphic unitary time evolution, the Hamiltonian part $H$
of the Liouville operator is not directly involved in the process
of transfer of information from the quantum subsystem to the
classical one. Only the dissipative part can achieve such a transfer
in a finite time.
\vskip 1truecm
\noindent{\underbar{\bf 2.2 The quantum mechanical measurement process}}
\medskip
In [3] we propose a simple, purely dissipative Liouville operator
(i.e. we put $H = 0$) that describes an interaction of $\sum_q$
and $\sum_c$, for which $m = n +1$ and $V_i = e_i \otimes \phi_i$,
where $\phi_i$ is the flip transformation of $X_c$ transposing the
neutral state $s_0$ with $s_i$. We show that the Liouville equation
can be solved explicitly for any initial state $\rho (0)$ of the
total system. Assume now that we are able to prepare at time
$t = 0$ the initial state of the total system $\Sigma_{tot}$ as an
uncorrelated product state $\rho(0) = w \otimes P^\epsilon (0),
P^\epsilon (0) = (p^\epsilon_0, p^\epsilon_1, ..., p^\epsilon_n)$
as initial state of the classical system parametrized by
$\epsilon, 0 \le \epsilon \le 1$:
$$
p^\epsilon_0 = 1 - {{n \epsilon} \over {n+1}},\eqno(8)
$$
$$
p^\epsilon_1 =  {\epsilon \over {n+1}}.\eqno(9)
$$
In other words for $\epsilon = 0$ the classical system starts from
the pure state $P(0) = (1,0,...,0)$ while for $\epsilon = 1$
it starts from the state $P^\prime (0) = ( {1 \over {n+1}},
{1 \over {n+1}}, ..., {1 \over {n+1}})$ of maximal entropy. Computing
$p_i(t) = Tr (\rho_i (t))$ and then the normalized distribution
$$
\tilde p_i (t) = {{p_i(t)} \over {\sum^n_{r=1} p_\tau (t)}}\eqno(10)
$$
with $\rho (t) = (\rho_0 (t), \rho_1 (t), ..., \rho_n (t))$ the
state of the total system we get:
$$
\tilde p_i (t) = q_i + {{\epsilon (1-nq_i)} \over {\epsilon n +
{{(1-\epsilon)(n+1)} \over 2} (1-e^{-2\kappa t})}},\eqno(11)
$$
where we introduced the notation
$$
q_i = Tr (e_i w),\eqno(12)
$$
for the initial quantum probabilities to be measured. For
$\epsilon = 0$ we have $\tilde p_i (t) = q_i$ for all $t > 0$,
which means that the quantum probabilities are exactly, and
immediately after switching on of the interaction, mirrored by
the state of the classical system. For $\epsilon = 1$ we get
$\tilde p_i (t) = 1/n$. The projected classical state is still
the state of maximal entropy and in this case we get no
information at all about the quantum state by recording the time
evolution of the classical one. In the intermediate regime, for
$0 < \epsilon < 1$, it is easy to show that $\vert \tilde p_i (t)
- q_i \vert$ decreases at least as $2 \epsilon (1 + e^{-2\kappa t})$
with
$\epsilon \to 0$ and $t \to + \infty$. For $\epsilon = 0$, that
is when the measurement is exact, we get for the partial quantum
state
$$
\pi_q (\rho (t)) = \sum_i e_i we_i + e^{-\kappa t} (w - \sum_i e_i we_i),
$$
so that
$$
\pi_q (\rho (\infty)) = \sum_i e_iwe_i,\eqno(13)
$$
which means that the partial state of the quantum subsystem
$\pi_q (\rho (t))$ tends for $\kappa t \gg + \infty$ to a limit which
coincides with the standard von Neumann-L\"uders quantum measurement
projection postulate.\hfill\break
\leftline{\bf Remark:}\hfill\break
The normalized distribution $\tilde p_i (t)$ is nothing else as
the read off from the outputs $s_1 ... s_n$ of the classical system
$\sum_c$.
\vskip 1 truecm
\noindent\underbar{{\bf 2.3 Efficiency versus accuracy by measurement}}
\medskip
Let us consider the case where
$$
V_i = \sqrt \kappa e_i \otimes f_i,\eqno(14)
$$
$f_i$ being the transformation of $X_c$ mapping $s_0$ into $s_i$.
In the Liouville equation we consider also an Hamiltonian part. We
find for the Liouville equation:
$$
\eqalignno{
\dot \rho_0 &= - i [H, \rho_0] - \kappa \rho_0, &(15)\cr
\dot \rho_i &= - i [H, \rho_i] + \kappa e_i\rho_0 e_i, &(16)\cr}
$$
where we allow for time dependence i.e. $H = H(t), e_i = e_i (t)$.
Setting $r_0(t) = Tr (\rho_0(t)),$\hfill\break
$r_i (t) = Tr (\rho_i (t))$, and
assuming that the initial state is of the form $\rho = (\rho_0, 0,
..., 0)$ we conclude that $\dot r_0 = - \kappa r_0$ and thus $r_0(t) =
e^{-\kappa t}$ which  implies that
$$
\sum^n_{i=1} r_i(t) = 1 - e^{-\kappa t} ,\eqno(17)
$$
from which it follows that a 50 \% efficiency requires $\log 2/\kappa$
time of recording. It is easy to compute $r_i(t)$ and
$$
\tilde p_i (t) = {{r_i(t)} \over {\sum^n_{j=1} r_j (t)}}
$$
for small $t$. We  get
$$
\tilde p_i (t) = q_i + {{\kappa^2t^2} \over 2} {1 \over \kappa}
\langle {{de_i} \over {dt}} \rangle_{\rho_0} +o(t^2),\eqno(18)
$$
where
$$
{{de_i} \over {dt}} \dot = {{\partial e_i} \over {\partial t}}
+ i [H, e_i].\eqno(19)
$$
Efficiency requires $\kappa t >> 1$ while accuracy is achieved if
$(\kappa t)^2 << {\kappa \over {\langle \dot e_i\rangle_{\rho_0}}}$.
To monitor effectively and accurately non stationary processes we must
therefore take  $t << \langle \dot e_i
\rangle_{\rho_0}$ and $1/\kappa << t$.
If however $H$  and $e_i$ does not depend on
time, then if either $\rho_0 (0)$ or $e_i$
commutes with $H$, we get $\tilde p_i (t) = q_i$ exactly and instantly.

When discussing measurement problem we may also ask the question
about the value of disturbance of the initial quantum state owing
to the measurement. Here we may look at the Eqns. (15-16) and
compare them to the undisturbed evolution ($\kappa = 0$) . Denoting
by ${\hat \rho} = \sum_\alpha \rho_\alpha$ the partial state of the quantum
system, we get for the difference of rates the two evolutions
$$\delta {\dot {\hat \rho}} = \kappa (\sum_i e_i \rho_0 e_i -
\rho_0 ) . $$
But for a generic initial state $\rho_0 $, the quantity
$\sum_i e_i \rho_0 e_i - \rho_0 $ can have trace norm of order $1$.
It follows that the disturbance of the state during a short
coupling time $t$ can be of the order $\kappa t$. Comparing this
to the discussion of efficiency, we conclude that efficiency of
the counters and their nondemolition property can be considered
in the model that we have discussed as being complementary to
each other.
\vskip 1 truecm
\noindent\underbar{{\bf 2.4 Stern Gerlach experiment}}\medskip
In the spirit of A. B\"ohm (cf. Ref. [8, Ch. XIII]) we model a
Stern-Gerlach device by a pure spin 1/2 particle interacting with a
spinless atom. Assuming that the magnetic field is linear in $z$
the interaction Hamiltonian can be written
$$
H_{int} = 2 \mu_B B z \sigma_3.\eqno(20)
$$
Writing
$$
\sigma_3 = {{\vert \uparrow \rangle \langle \uparrow \vert -
\vert \downarrow \rangle \langle \downarrow \vert} \over 2},
$$
$H_{int}$ is now given by
$$
H_{int} = \mu_B B (\vert \uparrow \rangle \langle \uparrow \vert
z + \vert \downarrow \rangle \langle \downarrow \vert (-z)).\eqno(21)
$$
Supposing now that the atom can be directly observed we can replace
it for all practical purposes by a 3-state classical device
$(s_0, s_+, s_-)$. The coupling is then modelled by
$$
\sqrt \kappa (p flip (0 \to +) + (1 - p) flip (0 \to -))\eqno(22)
$$
where $p = {1\over 2} (\sigma_0 + {\underline n} {\underline\sigma})$
is the projection onto the spin component to be measured.
We are now in position to approximate Stern-Gerlach experiment
by our 3-state model. For more details see also [6].
\vskip 1 truecm
\noindent\underbar{{\bf 2.5 Model of a counter}}\medskip
We consider as in [9] a one-dimensional ultra-relativistic quantum
mechanical particle. This nice model can be solved exactly and
provides a clear understanding of the physical phenomena at work.
The counter sensitivity is described by an operator valued function
$f(t)$ and the quantum system $\sum_q$ with ${\cal H}_q =
L^2 ({\bf R}, dx)$ is coupled to a 2-state classical system. The
Liouville equation for the state of the total system is
$$
\dot \rho = - i [H, \rho] + V \rho V^* - {1 \over 2} \{V^*V, \rho \},
\eqno(23)
$$
with
$$
V = f \otimes \left( \matrix{ 0 &1 \cr
                              0 &0 \cr}\right)
= \left( \matrix{ 0 &f \cr
                  0 &0 \cr}\right)\eqno(24)
$$
Now explicitly we obtain
$$
\dot \rho_0 = -i [H, \rho_0] - {1 \over 2} \{f^*f, \rho_0 \},\eqno(25)
$$
$$
\dot \rho_1 = -i [H, \rho_1] + f \rho_0f^*.\eqno(26)
$$
Taking $H = {1 \over i} {d \over {dx}}$ and $f = f^* = f(x,t)$
we obtain for the counting rate $\dot \rho_1 (t)$ in a free evolving
state:
$$
\dot p_1 (t) = \int_{\bf R} \vert \Psi (x-t) \vert^2 f^2 (x,t)
e^{- \int^t_0 f^2 (x+s-t,s)ds} dx.\eqno(27)
$$
Assume now that we have to do with a point article i.e.
$$
\vert \Psi (x) \vert^2 = \delta (x - x_0)
$$
we obtain in this idealized case
$$
\dot p_1 (t) = f^2 (x_0 + t,t) e^{- \int^t_0 f^2 (x_0+s,s)ds}\eqno(28)
$$
which expresses the fact that the counting rate depends on how
long the ``detector'' was already in contact with the particle. It is
possible to generalize the results obtained to the case of a
nonrelativistic quantum mechanical particle for which
$H = - {{d^2} \over {dx^2}}$. In this situation the model would remain
solvable within reasonable approximations. For more details see [6].
Moreover for the ultrarelativistic dynamics the multidetector case
can be solved completely.
\vskip 1.5 truecm
\noindent\underbar{\bf 3. Quantum Zeno Effect}\medskip
For a rapid sequence of measurements made at times $k \tau$ which
are multiples of a small unit time $\tau$, a quantum system will not
change at all. It seems as if time has stood still for it. This
effect known as Zeno effect can be understood in terms of quantum
mechanical perturbation theory. Indeed it gives for small times
a transition probability per unit of time
$$
p_1 (\tau) = 1 - \vert < \psi_0, e^{i \tau H} \psi_0> \vert^2
\simeq ( \Delta H )^2 \tau^2
$$
with
$$
(\Delta H)^2 = \langle \psi_0 , H^2 \psi_0 \rangle - \mid \langle
\psi_0 , H \psi_0 \rangle\mid^2 \geq 0
$$
and therefore $p_1 (\tau)$ vanishes
quadratically in the limit $\tau \to 0$.
For larger values of time the expected constant rate is formal. This
result must be explained if Quantum Theory is to continue to make
sense. The most obvious explanation is that any actual measurement
requires a time $T$ and the paradox is eliminated if it can be
shown that $T > \tau$.
\par
This effect was formulated by Turing 1940 and called 1977 Quantum Zeno
effect by Misra and Sudarshan [10]. In recent years there has been
considerable discussion of the quantum Zeno process, effect and
paradox.
See for example [11, 12, 13, 14]. Moreover it has been claimed that
experiments can demonstrate the effect [15, 16, 17].
\vskip 1,5truecm
\noindent\underbar{\bf 3.1 Quantum Zeno Effect revisited}
\medskip
Using our model
of a continuous measurement we can easily discuss this effect for a
quantum spin 1/2 system coupled to a 2-state classical system [18].
We consider only one orthogonal projector $e = e^* = e^2$ on the
two-dimensional Hilbert space ${\cal H}_q = {\bf C}^2$.
\par
To define the dynamics we choose the coupling operator $V$ in the
following way:
$$
V = \sqrt\kappa \left( \matrix {0, & e \cr
                                e, & 0 \cr} \right).
\eqno(29)
$$
The Liouville equation (7) for the density matrix $\rho = diag (\rho_0,
\rho_1)$ of the total system reads now
$$
\eqalign{
\dot\rho_0 &= - i [H, \rho_0] + \kappa (e\rho_1 e - {1\over 2}
\{ e, \rho_0\} ),  \cr
\dot\rho_1 &= -i [H, \rho_1] + \kappa (e\rho_0 e - {1\over 2}
\{ e, \rho_1 \} ).\cr}
\eqno(30)
$$
For this particularly simple coupling the effective quantum
state $\hat\rho = \pi_q (\rho) = \rho_0 + \rho_1$ evolves
independently of the state of the classical system, expressing
the fact that here we have only transport of information from the
quantum system to the classical one. We have:
$$
\dot{\hat\rho} = - i [H, \hat\rho] + \kappa (e\hat\rho e - {1\over 2}
\{ e , \hat\rho \} ).
\eqno(31)
$$
For the discussion of the quantum Zeno effect we specialize:
$$
\eqalign{
H &= {\omega\over 2} \sigma_3, \cr
e &= {1\over 2} (\sigma_0 + \sigma_1 ), \cr}
\eqno(32)
$$
$\sigma_i$ being the Pauli matrices, $i = 0, 1, 2, 3$.
\par
We start with the quantum system being initially in the
eigenstate of $\sigma_1$, and repeatedly (with "frequency" $\kappa$)
check if the system is still in this state, each "yes" causing a flip
in the coupled classical device - which we can continuously observe.
\par
The evolution equation for $\hat\rho$, with the initial condition
$\hat\rho (t = 0 ) = e$, can be exactly solved with the result:
$$
\hat\rho (t) = {1\over 2} (\sigma_0 + x (t) \sigma_1 + y (t)
\sigma_2 ),
\eqno(32)
$$
where $x(t), y(t)$ are given by
$$
\eqalign{
x(t) &= \exp \left( {{-\kappa t}\over 4} \right)
\left( {\rm cosh} \left( {{\kappa_\omega t}\over 4}\right) + {\kappa
\over \kappa_\omega} {\rm sinh} \left( {{\kappa_\omega t}\over 4}
\right)\right), \cr
y(t) &= {{4\omega}\over\kappa_\omega} \exp \left( {{-\kappa t}\over 4}
\right){\rm sinh} \left( {{\kappa_\omega t}\over 4} \right) , \cr}
\eqno(33)
$$
where $\kappa_\omega = \sqrt{\kappa^2 - 16\omega^2}$. Let us
introduce the dimensionless characteristic coefficient $\alpha =
{\kappa\over {4\omega}}$. For $\alpha > 1$ oscillations are damped
completely, and then the distance travelled by the quantum
state during the interaction becomes inversely proportional to the
square root of $\alpha$. The natural distance in the state space is the
geodesic Bures-Uhlmann distance $d_\frown$, which is the geodesic
distance for the Riemannian metric - given in our case by $ds^2 = g_
{ij} dx^i dx^j$, with $g_{ij} ( {\bf v} ) = (\delta_{ij} +
v_i v_j / (1 - {\bf v}^2))$. For density matrices $v = ( \sigma_0 +
{\bf v} \cdot \sigma)/ 2$ and $\omega = (\sigma_0 + {\bf w} \cdot
\sigma )/ 2$ we obtain
$$
d ( v \frown w ) = {1\over 2} {\rm arccos} ({\bf v} \cdot {\bf w}
+ ^4\sqrt{1 - {\bf v}^2} ^4\sqrt{1 - {\bf w}^2} ).
\eqno(34)
$$
In particular, if one of the states, say $v$, is pure, then
${\bf v}^2 = 1$ and $d (v\frown w )$ is simply given by
$$
d ( v \frown w ) = {1\over 2} {\rm arccos} ({\bf v} \cdot {\bf w} ).
\eqno(35)
$$
For $v = \hat\rho (t), w = e = (\sigma_0 + \sigma_1 )/2$, as in the
Zeno model, we obtain
$$ d (\hat\rho (t) \frown e ) = {1\over 2} {\rm arccos} (x (t)).
\eqno(36)
$$
Notice that $e$, being a pure state, is on the boundary of the
state space, and the $d_\frown$-distance from $e$ depends only on one
of the two relevant coordinates $x, y$ - contrary to the Frobenius
distance $T_r((v-w)^2)$, which would involve both coordinates.
Assuming now $\alpha \gg 1$ and $\kappa t \gg 1$, we get for
$x(t), y(t)$ in (33) asymptotic formulae:
$$
\eqalign{
x(t) &\asymp 1 - {{2\omega^2 t}\over\kappa} + \ldots \cr
y(t) &\asymp {\omega\over {2\kappa}} + \ldots \cr}
\eqno(37)
$$
Thus the distance reached by state is in this asymptotic region given
by
$$
d\asymp \omega \sqrt{{t\over \kappa}}
\eqno(38)
$$
\vfill\eject
\vskip 1.5truecm
\noindent\underbar{\bf 3.2 Quantum Zeno effect via piecewise
deterministic processes}
\medskip
Our objective in this section is to give a stochastic description of the
continuous measurement
Zeno effect by using piecewise deterministic processes. A noteworthy
reference for this subject is [19].
An observable $A$ of the total system is a pair $(A_\alpha )_{\alpha
=0,1}$ of operators in ${\cal H}_q$. Every observable $A$ determines a
function $f_A$ on pure states of the total system (or a pair of
functions $f^\alpha_A$ on pure states of $\Sigma_q$) by
$$
f_A (\varphi , s_\alpha ) \equiv f^\alpha_A (\varphi ) \equiv
\langle\varphi , A_\alpha \varphi \rangle .
\eqno(38)
$$
We shall exhibit a Markov semigroup generator, $M$, acting on
functions $f (\varphi , s)$ which gives the evolution equations (30)
in their dual form:
$$
\dot A_\alpha = i [H,A_\alpha] + \kappa (e A_\alpha e - {1\over 2}
\{ e,A_{\alpha^\prime} \} )
\eqno(40)
$$
where we use the notation $\alpha^\prime = \alpha + 1$ mod 2. Thus we
want to have
$$
\eqalign{
(M f_A ) ( \varphi , s_\alpha ) &=
{d\over{dt}} f_{A (t)} (\varphi , s_\alpha ) \mid_{t=0} \cr
&= {d\over{dt}} \langle \varphi , A_\alpha (t) \varphi \rangle \mid_
{t=0} = \langle \varphi , \dot A_\alpha \varphi \rangle .\cr}
\eqno(41)
$$
To this end we have to rewrite the RHS of (41)
$$
\langle \varphi , \dot A_\alpha \varphi \rangle = \langle \varphi ,
i[H , A_\alpha ]\varphi\rangle + \kappa \langle\varphi , eA_\alpha
\varphi \rangle - {\kappa\over 2} \langle \varphi , \{ e,
A_{\alpha^\prime} \} \varphi \rangle
\eqno(42)
$$
in terms of functions $f^\alpha_A (\varphi ) = f_A (\varphi , s_\alpha
)$. Note that the first term on the RHS of (42) is equal to
$$
\langle\varphi , i [H , A_\alpha ]\varphi \rangle = {d\over {dt}}
f^\alpha_A (e^{-iHt}\varphi ) \mid_{t=0} .
\eqno(43)
$$
By introducing the vector field $X_H$ on the unit ball of ${\cal H}_q$
defined by the Hamiltonian evolution:
$$
(X_H f ) (\varphi) \dot= {d\over{dt}} f ( e^{-iHt} \varphi ) \mid_{
t=0}
\eqno(44)
$$
we observe that
$$
\langle \varphi , i[H , A_\alpha ] \varphi \rangle = (X_H f^\alpha_A )
(\varphi ) .
\eqno(45)
$$
The next step is to observe that
$$
- {\kappa\over 2} \langle\varphi , \{ e , A_\alpha \} \varphi \rangle
= {d\over{dt}} \langle e^{-{{\kappa t}\over2} e} \varphi ,
A_\alpha e^{-{{\kappa t}\over 2} e} \varphi \rangle \mid_{t=0} ,
\eqno(46)
$$
which will give rise to two terms as follows. Let us introduce another
vector field $X_{\cal D}$ on the unit ball of ${\cal H}_q$ by
$$
(X_{\cal D} f ) (\varphi ) \dot= {d\over{dt}} f \left( {{e^{- {{\kappa t}
\over 2}e} \varphi}\over {\parallel e^{-{{\kappa t}\over 2}e}
\varphi \parallel}} \right) \mid_{t=0} .
\eqno(47)
$$
We are now able to rewrite (46) as
$$
- {\kappa\over 2} \langle \varphi , \{ e , A_\alpha \} \varphi \rangle
= (X_{\cal D} f^\alpha_A ) (\varphi ) - \lambda (\varphi ) f^\alpha_A
(\varphi ) ,
\eqno(48)
$$
where we introduced the function $\lambda (\varphi )$
$$
\lambda (\varphi ) = \kappa \parallel e\varphi \parallel^2 .
\eqno(49)
$$
We can express now the middle term of (42) as
$$
\kappa \langle \varphi , e A_{\alpha^\prime} e\varphi \rangle = \lambda
(\varphi) f^{\alpha^\prime}_A \left( {{e\varphi}\over{\parallel
e\varphi \parallel}} \right) .
$$
It follows that
$$
\langle \varphi , \dot A_\alpha\varphi\rangle = (X_H + X_{\cal D})
f^\alpha_A
(\varphi) +\lambda (\varphi ) A_{\alpha^\prime} \left( {{e\varphi}
\over{\parallel e\varphi\parallel}} \right) -\lambda (\varphi ) A_
\alpha (\varphi ) .
\eqno(50)
$$
We finally introduce the matrix valued measure
$$
Q ( d\psi ,\varphi ) = (Q_\alpha^\beta (d\psi , \varphi )) =
\left( \matrix{ 0 , & \delta_{{\varphi\over{\parallel e\varphi
\parallel}}} \cr
\delta_{{\varphi\over{\parallel e\varphi \parallel}}} , & 0 \cr}
\right) ,
\eqno(51)
$$
then
$$
\eqalign{
\langle \varphi , \dot A_\alpha \varphi \rangle &=
(X_H +X_{\cal D} ) f^\alpha_A (\varphi ) \cr
&+ \lambda (\varphi) \sum_\beta \int (f^\beta_A (\psi) - f^\alpha_A
(\varphi )) Q_\beta^\alpha (d\psi , \varphi ) . \cr}
\eqno(52)
$$
It shows that the evolution equation (40) follows from a Markov
semigroup of a piecewise deterministic process with generator
$$
\eqalign{
(M, f) (\varphi , s_\alpha ) &=
[( X_H + X_{\cal D} ) f] (\varphi , s_\alpha ) \cr
&+ \lambda (\varphi ) \sum_\beta \int [ f (\psi , \beta ) - f (\varphi
,\alpha )] Q_\beta^\alpha (d\psi , \varphi ) . \cr}
\eqno(53)
$$
To this process, as in [19], we can associate a jump process which
is not a Markov
process and a Feynman-Kac formula can be used to calculate the
expectations of functionals. We refer to [20] for technicalities.
The information contained in the Liouville equation (39) is
therefore not the maximal available one. Knowing the piecewise
deterministic Markov process associated to (53) we want to emphasize
that we can answer all kinds of questions about time correlations
of events and also simulate the random behavior of the classical
system $\Sigma_c$ coupled to the quantum system $\Sigma_q$.
Let $T_t$ be the one parameter semigroup of non-linear transformations
of rays in ${\bf C}^2$ given by
$$
T_t \phi = {{\phi (t)}\over{\parallel \phi (t) \parallel}}
\eqno(54)
$$
where
$$
\phi (t) = e^{-iHt -{k\over2} et} \phi
\eqno(55)
$$
Let us now suppose that a $t=0$ the quantum system $\Sigma_q$
is in the pure state $\varphi$ and the classical system in the
state $s_\alpha$. Then $\varphi$ starts to evolve according to
the deterministic non unitary (and non-linear) time evaluation
$T_t \varphi$ until jump occurs at time $t_1 > 0$. The random
jump-time $t_1$ is governed by an inhomogeneous Poisson
process with rate function
$$
\lambda (t) = \kappa \parallel e T_t \varphi \parallel^2 .
\eqno(56)
$$
The classical system switches from $s_\alpha$ to $s_{\alpha^\prime}$,
while $T_{t_1}\varphi$ jumps to $eT_{t_1}\varphi / \parallel e T_{t_1}
\varphi\parallel$ with probability 1, and the process starts again.
If the initial state $\varphi$ is an eigenstate of $e$ with
eigenvalue one
$$
e\varphi = \varphi
\eqno(57)
$$
as it is in our Zeno model, and for large values of the coupling
constant $\kappa$, the intensity $\lambda$ is nearly constant and
equal to $\kappa$.
Thus $1/\kappa$ can be interpreted as the mean time between
succesive jumps. Strong coupling between the classical and the
quantum system, which is necessary for the occuring of the Zeno
effect, manifests itself by a high frequency of jumps.
Notice that the distribution function of the jump time is given by
$$
p[t_1 > t ] = \exp ( - \int^t_0 \lambda ( T_s\varphi ) ds ) ,
\eqno(58)
$$
and so the probability that the jump will occur in the time interval
$(t, t+dt)$ is
$$
- {{dp [t_1 > t]}\over{dt}}
= \parallel e T_t \varphi \parallel^2 \exp \left( -
\int^t_0 \parallel e T_s \varphi \parallel^2 ds \right).
\eqno(59)
$$
Thus at time instants $t$ when $\parallel e T_t \varphi \parallel =0$,
which would cause problems with the formulae (50) and (51), jumps do not
occur. In section 4 we will comment on the meaning of these jumps.
\par
We notice that the conventional analysis of the Zeno effect, as given
in the analysis of experiment by Itano (see Refs [15], [16]) is in
agreement with our framework.
\vskip 1.5truecm\noindent
\underbar{\bf 3.3 Remarks on "meaning of the wave function"}
\medskip
It is tempting to use the Zeno effect for slowing down the time
evolution in such a way, that the state of a quantum system $\Sigma_q$
can be determined by carrying out measurements of sufficiently many
observables. This idea, however, would not work, similarly like would
not work the proposal of "protective measurements" of
Y. Aharonov et al (see [21] [22]). To apply Zeno-type measurements just as
to apply a "protective measurement" one would have to know the state
beforhand. Also, our discussion in Sec 2.3 suggests that obtaining
a reliable knowledge of the quantum state may necessarily lead to
a significant, irreversible disturbance of the state.
This negative statement does not mean that we have shown that
the quantum
state cannot be objectively determined. We believe however that
dynamical, statistical and information-theoretical aspects
of the important problem of obtaining a "{\sl maximal reliable knowledge
;, of the unknown
quantum state with a least possible disturbance}"
are not yet sufficiently understood.
\vskip 1.5 truecm
\noindent\underbar{\bf 4. Concluding remarks and comments}\medskip
One aim of this review was to show how the problem of quantum
measurements can be tackled by using consistent models of interactions
between classical and quantum systems. We believe that the framework
we propose is not only in position of describing theoretically the
measurement process but is also able of analyzing correctly recent
experiments. We described models providing an answer to the problem
of how and when a quantum phenomenon becomes real. The central idea of
these models is based on a modification of quantum mechanics by
introducing dissipative elements in the basic dynamical equation and
on allowing for a nontrivial dynamics of central quantities
\par
The minimal piecewise deterministic process introduced in
connection with Zeno effect can be used for computing time
characteristic of the interaction and also for numerical
simulations of the phenomenon. One may ask "are these jumps
\underbar{real}?". Our answer is: yes, they are \underbar{real} -
to the extent that they can be such. These jumps do not occur in space
and time but they occur in ${\cal H}_q$. Let us emphasize that they
can be detected all the same by monitoring devices that are placed in
space and time. Our formalism is on this respect consistent:
indeed we give not only a theory of jumps but also give means of their
"experimental" detection. The complete theory of monitoring of
quantum systems, including the analysis of disturbance and
information transfer is still to be worked out.

\par
Bohr and Heisenberg made avery sharp distinction between the
classical and the quantum domains. The borderline seemed to
coincide with the division between macroscopic and microscopic.
With the Josephson junctions we are obliged to accept the
existence of macroscopic quantum mechanical systems. In [4] and [23]
we show that our framework is very well adapted to give a
description of a "mini" SQUID and to discuss the behaviour of the
coupled system consisting of a macroscopic classical system
(tank circuit) and a single quantum object (SQUID) [cf. Ref 24].
The same method
can be applied as well to other problems where the classical
system is expected to respond to averages of some quantum
observables, like, for instance, classical gravitational field is
expected to have as its source averaged energy-momentum tensor of
quantized matter.
\par
Science continually reworks its foundation and even the formalism of
Quantum Mechanics, in spite of the fact that it fits Nature like
a glove is not immune to change.
The problem of quantum mechanical measurement has been solved
"in practice". Indeed Quantum Mechanics is used daily and it works.
Our aim was to analyze and to understand how and why it works. Our
results show that the reduction of the wave function is not an
added postulate but a necessary consequence of the time evolution
of the total system $\Sigma_t = \Sigma_c \times \Sigma_q$. The
classical probabilities we obtain depend on the available
knowledge about the system, giving us the best predictions possible
from the partial information that we have.
\par
Einstein once wrote to Schr\"odinger that "the Heisenberg-Bohr
tranquilizing philosophy is so delicately contrived that, for the time
being, it provides a gentle pillow for the true "believer". The
pillow we propose does not aspire to be a miraculous youth
elixir for Quantum Theory making it universal and true for ever.
But perhaps
it will be fatter and firmer and will help to stop the bleeding
from some open scars ... .
\vskip 1.5truecm\noindent
{\bf Acknowledgements}
\medskip\noindent
The support of Alexander von Humboldt Foundation is acknowledged with
thanks.
\vfill\eject
\noindent\underbar{\bf References}\medskip
\item{[1]} N. Bohr, a) Atomic Theory and the Description of Nature,
Cambridge University Press (1934)\hfill\break
b) Atom Physics and Human Knowledge, Wiley (1963)
\item{[2]} F. Rohrlich, From Paradox to Reality, Our basic concepts
of the physical world, Cambridge University Press (1987)
\item{[3]} Ph. Blanchard, A. Jadczyk, On the interaction between
classical and quantum systems, Physics Letters \underbar{A175} (1993)
157-164
\item{[4]} Ph. Blanchard, A. Jadczyk, Classical and Quantum Intertwine,
in {\it Proceedings of the Symposium of Foundations of Modern Physics},
Cologne, June 1993, Ed. P. Mittelstaedt, World Scientific (1993)
\item{[5]} R. Haag, Irreversibility introduced on a fundamental level,
Comm. Math. Phys. \underbar{123} (1990) 245-251
\item{[6]} Ph. Blanchard, A. Jadczyk, Coupled quantum and classical
systems, measurement process, Zeno's effect and all that, in
preparation
\item{[7]} Ph. Blanchard, A. Jadczyk, Nonlinear effects in coupling
between classical and quantum systems: SQUID coupled to a classical
damped operator, to appear
\item{[8]} A. B\"ohm, Quantum Mechanics, Springer Verlag (1979)
\item{[9]} H. Nakazato, S. Pascazio, Solvable Dynamical Model for a
Quantum Measurement Process, Phys. Rev. Lett. \underbar{70} (1993)
1-4
\item{[10]} B. Misra and E.C.G. Sudarshan, The Zeno's paradox in
quantum theory, J. Math. Phys. \underbar{18} (1977) 756-763
\item{[11]} L. Accardi, The probabilistic roots of the
Quantum Mechanical Paradoxes, in the "Wave - Particle Dualism",
S. Diner (Ed.) 294-330, S. Reidel (1984)
\item{[12]} D. Home and M.A.B. Whitaker, A critical re-examination of
the quantum Zeno paradox, J. Phys. \underbar{A25} (1992) 657-664
\item{[13]} E. Joos, Continous measurement: Watchdog effect versus
golden role, Phys. Rev. \underbar{D29} (1984) 1626-1633
\item{[14]} H. Fearn and W.E. Lamb, Jr. Computational approach to the
quantum Zeno effect. Position measurements, Phys. Rev.
\underbar{A46} (1992) 1199-1205
\item{[15]} R.J. Cook, What are Quantum Jumps, Phys. Scr. \underbar{
T21} (1988) 49-51
\item{[16]} W.M. Itano, D.J. Heinzen, J.J. Bollinger and
D.J. Wineland, Quantum Zeno effect, Phys. Rev. \underbar{A41} (1990)
2295-2300
\item{[17]} S. Pascazio, M. Namiki, G. Badurek and H. Rauch, Quantum
Zeno effect with neutron spin, Physics Letters \underbar{A179}
(1993) 155-160
\item{[18]} Ph. Blanchard and A. Jadczyk, strongly coupled quantum and
classical systems and Zeno's effect, to appear in Physics Letters A
\item{[19]} M.H.A. Davis, Lectures on Stochastic Control and
Nonlinear Filtering, Tata Institute of Fundamental Research,
Bombay (1984)
\item{[20]} Ph. Blanchard, A. Jadczyk, On random processes
associated with Quanto-Classical Interactions, in preparation
\item{[21]} Y. Aharonov, J. Anandan and L. Vaidman, Meaning of the
wave function, Phys. Rev. A47 (1993) 4616-4626
\item{[22]} Y. Aharonov and L. Vaidman, Measurement of the
Schr\"odinger wave of a single particle, Phys. Letters A178 (1993)
38-42
\item{[23]} Ph. Blanchard, A. Jadczyk, Nonlinear effects in
coupling between classical and quantum systems: SQUID coupled to a
classical damped oscillator, to appear
\item{[24]} T.P. Spiller, T.D. Clark, R.J. Prance and A. Widom,
Quantum Phenomena in Circuits at Low Temperature Physics, Vol XIII
(1992) 221-265
\end